# ACCELERATOR PHYSICS ISSUES IN THE BEPCII

L. M. Chen, Y. Luo, Q. Qin, J. Q. Wang, S. Wang, G. Xu, C. H. Yu, C. Zhang,

IHEP, Beijing 100039, China

*Abstract*

The Beijing Electron-Positron Collider (BEPC) has been running for both high energy physics (HEP) and synchrotron radiation (SR) researches since 1989. Good performance of BEPC accomplishes a lot of achievements in the τ-charm region over the past decade, and in many fields of SR applications as well. As an upgrading scheme from BEPC, the BEPCII project was approved by the Chinese government in the early time of this year, with a micro-β scheme and multi-bunch collision in a double-ring machine installed in the current BEPC tunnel. The design work of the BEPCII is being carried on. This paper will describe the main progresses on the design of the BEPCII in the field of accelerator physics.

## 1 BEAM-BEAM INTERACTION

To achieve the high luminosity in the factory class collider, high beam currents and small beam sizes are necessary. These induce a strong beam-beam interaction. The successful performance of KEKB and PEPII indicates that the beam-beam limit can be reached without any single bunch instability. This means the beam-beam interaction limits the peak luminosity. It is an important issue to study the beam-beam interaction in the design and performance of such a high luminosity collider.

The simulation studies are done by taking the advantages of the code BBC (Beam-Beam interaction with a Crossing angle) developed by K. Hirata [1]. BBC is a weak-strong simulation code in 6-D phase space, including the effect of crossing angle. Although the weak strong simulation can not investigate the coherent phenomena of beam-beam interaction, it is generally used during the design of a collider. Because of the CPU time consuming, the strong-strong simulation in a large scale tune scan so far has not yet been done.

The effect of a finite bunch length is taken into account by dividing a strong bunch into 5 slices longitudinally, and the weak one is represented by 50 randomly generated super particles, with a Gaussian distribution in 6-D phase space. The simulation is done for more than 5 radiation damping time.

The tune scan is performed for optimizing the tune from the viewpoint of high luminosity. Figure 1 shows the simulated luminosity on the tune grid (fractional part only) of $\delta\nu_x \in (0,1)$ and $\delta\nu_y \in (0,1)$ with a crossing angle of $\phi_c = 11$ mrad×2, in which the luminosity reduction factor $L/L_0$ is given instead of the luminosity itself. The mesh size is set as 0.02, which is smaller than the synchrotron tune $\nu_s = 0.034$. It indicates that the high luminosity region is just above the half integer in horizontal plane, and there is no significant difference between just above half integer and just above the integer in vertical plane. According to the commissioning experiences of KEKB and PEPII [2], a vertical tune above half integer is preferable because the orbit distortion is much stable than that of above an integer. The high luminosity is expected at around $\delta\nu_x=0.53$ and $\delta\nu_y=0.58$, and these tune values are set as the design working points.

To achieve higher luminosity, a larger beam-beam parameter $\xi_y$ is preferable. However, the maximum $\xi_y$ is limited by beam-beam interaction. The simulation results show that the maximum $\xi_y$ is decreased because of the large crossing angle of $\phi_c=11$ mrad×2, but the design value of $\xi_y = 0.04$ is reasonable and reachable.

For BEPCII, the crossing angle of $\phi_c=11$ mrad×2 is the basic requirements of the interaction region. From the viewpoint of beam-beam interaction, the crossing angle not only limits the maximum $\xi_y$, as described above, but also induces some additional luminosity reductions due to the geometric effects. The simulation shows that the luminosity reduction factor due to the finite bunch length effect and the crossing angle is about 80% while the luminosity reduction factor is 86% for the head-on collision. However, the crossing angle of $\phi_c = 11$ mrad×2 is still acceptable. The simulation also indicates that the bunch length should be controlled carefully to avoid further luminosity reduction due to the finite bunch length effect.

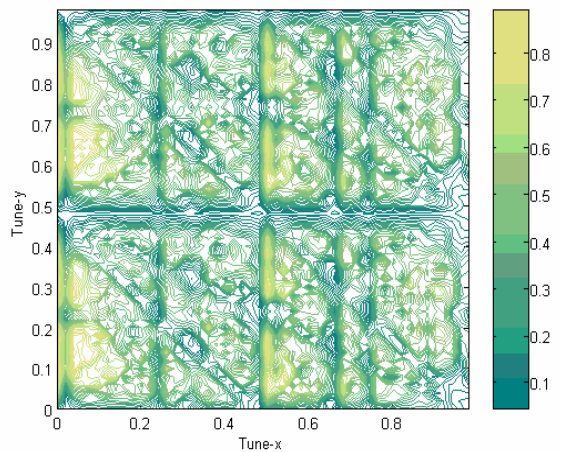

Figure 1: Luminosity survey with a crossing angle of $\phi_c = 11$ mrad×2

## 2 LATTICE DESIGN OF THE BEPCII STORAGE RING

With a new inner ring installed inside the old one in the existing BEPC tunnel, BEPCII will provide the colliding beams with the center-of-mass energy from 1.0 GeV×2 to 2.1 GeV×2 and also the dedicated synchrotron radiation beam at 2.5 GeV. For the colliding beams the luminosity is optimized at 1.89 GeV with the peak luminosity of $1\times10^{33}$ cm$^{-2}$s$^{-1}$. As to the dedicated synchrotron radiation mode, the beam current reaches 250 mA with an emittance as low as possible.

There have been several beam-lines in the BEPC for the synchrotron radiation experiments, so the current bending magnets and insertion devices in the southern region have to be fixed at their present positions. In order to increase the average luminosity, the "top-off" injection scheme up to 1.89 GeV is adopted. This requires that the injection and collision optics be the same. The main parameters of the storage ring for both collision and SR modes are shown in Table 1.

Table 1  Main Parameters of the BEPCII storage ring

| Parameters | Unit | Collision | SR |
|---|---|---|---|
| Energy | GeV | 1.89 | 2.5 |
| Circumference | m | 237.53 | 241.13 |
| RF frequency | MHz | 499.8 | 499.8 |
| Harmonic |  | 396 | 402 |
| RF voltage | MV | 1.5 | 3.0 |
| Transverse tunes |  | 6.53/7.58 6.53/5.58 | 8.28/5.18 |
| Damping time | ms | 25/25/12.5 | 12/12/6 |
| Beam current | A | 0.91 | 0.25 |
| Bunch number |  | 93 | Multi |
| SR loss per turn | keV | 121 | 336 |
| SR power | kW | 110 | 84 |
| Energy spread |  | $5.16\times10^{-4}$ | $6.66\times10^{-4}$ |
| Compact factor |  | 0.0235 | 0.016 |
| Bunch length | cm | 1.5 | 1.18 |
| Emittance | nm·rad | 144/2.2 | 120/ |
| $\beta$ function at IP | m | 1/0.015 | – |
| Crossing angle | mrad | 11× 2 | – |
| Bunch spacing | m | 2.4 | – |
| Beam-beam Parameter |  | 0.04/0.04 | – |
| Luminosity | cm$^{-2}$s$^{-1}$ | $1.0\times10^{33}$ | – |

*2.1 Geometric Design*

Being upgraded from BEPC, BEPCII will use the old tunnel and keep the present beam-lines, so that the circumferences of "three rings" (two horizontally separated colliding rings and an SR ring) and the distance between the outer and inner rings are only adjustable in a few tens of centimeters. The RF frequency of the colliding rings is constrained by the linac frequency and the demands of the harmonic number from the data acquisition system of the BESIII detector. The final choice of the RF frequency is 499.8 MHz, which is 7/40 of the linac frequency 2856 MHz. Corresponding to this frequency, the harmonic numbers for the colliding and SR rings are 396 and 402, respectively. The distance between the outer and inner rings is 1.18 m. The north half ring of BEPC must be shifted 0.36 m northward.

Due to the space limit, the only way to separate the two beams is to use the crossing angle at the interaction point (IP). The crossing angle of BEPCII is chosen as 11 mrad × 2. Since the crossing angle cannot give a sufficient separation, the first pair of defocusing quadrupoles of the IR will deflect the beams further to 26 mrad × 2 and then a pair of septum magnets deflects the beam in the inner rings to 65.5 mrad, avoiding the SR background in the IR in the meantime. A bridge is needed in the SR mode connect the two half outer rings. Two superconducting dipole coils on both sides of the IP are used to accomplish this function. The beam will have 6 mm horizontal offset at the bending coils for the SR mode. In the RF region, the crossing angle of the particle trajectories is 154.7 mrad×2. Vertical local bumps will be used to separate two beams, so that the optics of two rings can be symmetric. In the operation of the SR mode, the bridges, which connect the inner and outer rings, will be powered off. Thus, the electron beam will go straight through the bending magnets. In this case, both of the RF cavities can provide the power to the electron beam. Fig. 2 shows the BEPCII complexity.

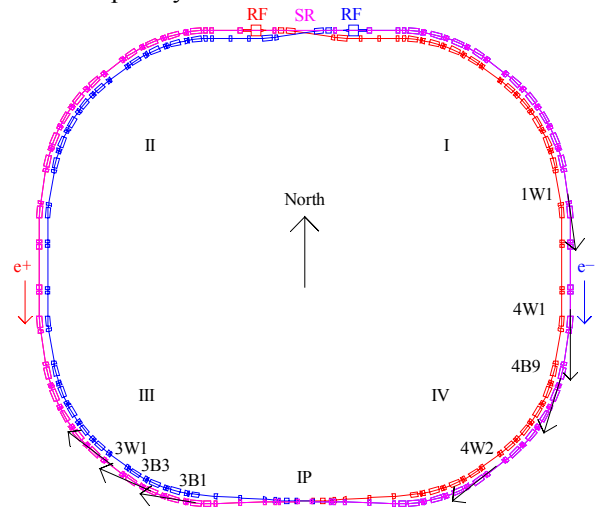

Figure 2: The BEPCII Complexity

*2.2 Interaction Region Design*

The design of the interaction region (IR) has to accommodate competing and conflicting requirements from the accelerator and the detector. Many equipment including magnets, beam diagnostic instruments, masks, vacuum pumps, and experiment detector must coexist in a very small region. So the most difficult part of the design of a new collider is that of the IR.

The IR of BEPCII is about ±14 m long from the IP, in which about 20 main magnets need to be installed. The space is very tight. The design criteria for the BEPCII IR configuration are based on the experiences from other machines and some special requirements of the BEPCII. The two beams collide at the IP with a horizontal crossing angle of 22 mrad. The IR configuration provides an effective compensation scheme of high field detector solenoid. The aperture of vacuum chamber in the IR must be designed at least 2×(14$\sigma$ +2 mm) for the beams (uncoupled in the horizontal plane and fully coupled in the vertical plane). And the background level of both synchrotron radiation and lost particles in the detector must be sufficiently low.

The BESIII detector consists of a cylindrical drift chamber, surrounded by an electromagnetic calorimeter, which is immersed in about 1 T magnetic field of a 3.7 m long superconducting solenoid. The geometry of the drift chamber requires that the accelerator components inside the detector should fit the conical space with an opening angle of 21.5˚. The first accelerator component, which follows the central drift chamber, can only approach to 0.552 m on each side of the IP. The superconducting magnet SCQ and the septum bend magnet ISPB are fully located inside the detector boundary. The schematic layout of the IR is shown in Fig 3.

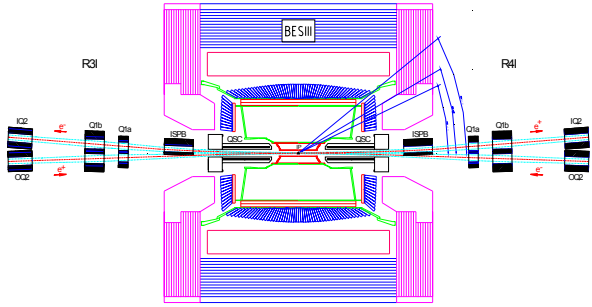

Figure 3: Schematic layout of the IR

In order to avoid the parasitic bunch crossing, a quick beam separation after collision is necessary. Our design choice is to introduce a horizontal crossing angle of 2×11 mrad at the IP and to incorporate with a horizontal bending magnet labeled ISPB for a larger separation between e− and e+ beams. The ISPB is a septum magnet with a narrow septum coil, located at 2.3 m away from the IP with a length of 0.6 m and the field strength of 4.6 kGs. It acts on the outgoing beam only.

On each side of the IP, a doublet of quadrupoles on each beam-line is used to provide the focusing optics needed at the IP ($\beta_y^*$=1.5 cm). The first vertical focusing quadrupole is a superconducting magnet (SCQ) shared by both beams. The SCQ package also includes compensation solenoid magnets, dipole magnets for SR mode and skew quadrupole coils. Its cryostat has a warm bore with an inner diameter of $\phi$134 mm and an outer diameter of $\phi$302 mm. The endcan of the cryostat has an outer diameter of $\phi$635 mm. The second elements of the doublets are horizontal focusing quadrupoles Q1a and Q1b with a distance of 0.3m between them. In order to keep the symmetry for the two rings and save the space, Q1a and Q1b are designed as a two-in-one structure and two separated beam channels for the incoming and outgoing beams with the same field strengths. These quadrupoles have to meet some special difficulties in design because the two beam pipes at their positions are still very close to each other. The next machine elements are quadrupoles Q2, Q3 and Q4, where the sufficient separation of the two beams is available to allow them to be installed side by side in the two rings. Those quadrupoles match the IR optics to that of the arc region.

### 2.3 Optics Design of the Collision Mode

The double-ring geometric structure of BEPCII causes each ring not to be a 4-fold symmetric structure, but the electron and positron rings are symmetric while the SR ring is east-west symmetric. Each ring for collision can be divided into four regions: IR, arc, injection, and RF regions.

In the IR region, two superconducting quadrupole (SCQ) are located at both sides of the IP to squeeze the $\beta_y^*$. It also bends the beams further from 11 mrad to 26 mrad. 5 warm bore quadrupoles are used for connecting the arc and IP. A low field bending magnet is located at the beginning of the arc to decrease the synchrotron radiation to the IP. Since the longitudinal space in the IR is very tight, it is difficult to adopt a compensation scheme using skew quadrupoles. Thus, the coupling is locally compensated inside the detector with anti-solenoids and skew quadrupole coils.

In the RF region, there are seven quadrupoles to connect the arc regions of the inner and outer rings. Part of the RF region is dispersion free, and the β functions at the RF cavities in both vertical and horizontal planes are less than 15 m.

At the injection points, dispersion is zero as well as the symmetry position in the inner ring. The horizontal β function is larger than 20 m to reduce the sigma amplitude of the remnant oscillation of the injected beam. The phase advance between two kickers is π.

The main consideration of the optics in the arc region is to meet the designed emittance, momentum compaction factor, tunes and a sufficient dynamic aperture. The β functions in both vertical and horizontal planes are less than 25 m, and the dispersion is less than 2.5 m. The horizontal β functions at focusing sextupoles and the vertical β functions at defocusing sextupoles are as large as possible. Moreover, the limited vertical aperture at the insertion devices leads to the constraint of vertical β function.

### 2.4 Chromaticity Correction and Dynamic Aperture

A sufficient dynamic aperture is necessary, either for efficient beam injection or long beam lifetime. Though the requirements for the injection and the colliding beams are slightly different, we use the condition of the injection beam for both cases, which means a large momentum acceptance (~10$\sigma_\varepsilon$) and large transverse apertures. For the tracking results evaluated here, the horizontal RMS beam size is taken from the natural horizontal emittance at 2.1 GeV (0.18 μm), and the vertical size $\sigma_y$ from the fully-coupled emittance, i.e., half of the uncoupled horizontal emittance. In the dynamic aperture tracking with SAD [3] program, particles are launched at the IP with 1000 turns.

36 sextupoles are used to correct natural chromaticities in each storage ring, powered by 18 power supplies. With these sextupoles, chromaticities of the BEPCII storage ring are corrected to 1.0.

Fig. 4 shows the tracking results for the BEPCII dynamic aperture. One can see that the dynamic aperture is significantly reduced if errors from magnets are included. Studies show that for 6.53/7.58 region the dynamic aperture is sensitive on $K_1$ errors for the dipoles and quadrupoles and more sensitive if the horizontal tune is closer to the half integer, this means the linear lattice must be well performed otherwise it will be difficult to operate the machine close to the half integer, for 6.53/5.58 region the situation is better than 6.53/7.58 region.

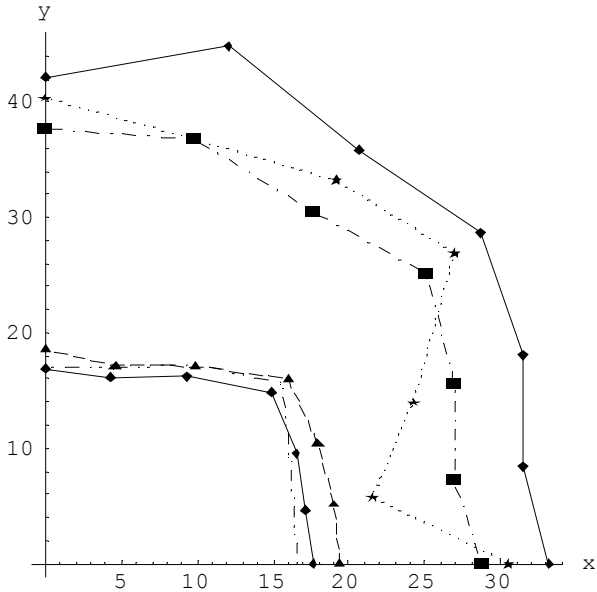

Figure 4: Dynamic aperture for BEPCII. The 3 most outer lines are for bare lattice: solid one is on-momentum, dotted and dot-dashed ones $\pm 10\sigma_\varepsilon$ off momentum. The 3 inner lines for the average of 20 random seeds with misalignment (after COD correction) and multiple errors (without $K_1$) : dashed line is on-momentum, solid and dot-dashed ones $\pm 10\sigma_\varepsilon$ off momentum.

## 2.5 Optics design of the SR mode

In the BEPCII complex, an electron ring called SR mode which can be used as a dedicated light source is formed by connecting the two outer half rings of the electron and positron rings. The design goal of the SR mode is to operate electron beam at 2.5 GeV with a maximum beam current of 250 mA. The betatron tunes are chosen as $v_x/v_y$ = 8.28/5.18. The emittance is thus 120 nm·rad at 2.5 GeV. If the dispersion at RF cavities is allowable, the emittance can reach 100 nm·rad or less. The maximum beta functions $\beta_x$ and $\beta_y$ are smaller than 23 m. The maximum horizontal dispersion function $D_x$ is about 1.65 m. The natural chromaticity is $\xi_{x0} = -11.6$ and $\xi_{y0} = -9.7$ for the horizontal and vertical planes, respectively. The dynamic aperture is larger than $25\sigma$ even with 0.8% energy spread.

## 2.6 Beam Injection

BEPCII will keep the BEPC's horizontal phase space multi-turn injection, and maintain its beam transport lines, which connect the linac and the ring. In order to achieve a high average luminosity, the beam injection rates have to be greatly improved, which are expected to be higher than 200 mA/min and 50 mA/min for e− and e+ beams, respectively. To ensure this high injection rates, for the positron beam, two-bunch injection scheme will be adopted with an adjustable injection repetition (12.5 Hz, 25 Hz, and 50 Hz).

The BEPC beam transport lines are upgraded in many aspects, such as replacing the old power supplies, adding four new independent power supplies to free some lattice matching constraints and installing new beam position monitors along the lines to keep close and real time control of the beam central orbits. In order to reduce the number of the particles lost in the ring, especially in the interaction region, collimators are needed in the beam transport lines to clean the particles with large betatron amplitude or energy deviation. Instead of the BEPC's injection through windows separating the transport line and the storage ring, the beam pipes of transport line will be directly connected to the storage ring to minimize the emittance dilution. So the vacuum condition in the end of the transport lines should be improved.

Two horizontal kickers with a phase advance of $\pi$ are employed to create the horizontal closed orbit bump for the circular beam. The Lamberston septa of BEPC are still kept. Since the bunch spacing is changed, new kicker magnet power supplies and vacuum chamber are being studied. The physical apertures given by the IR are $14\sigma_{x,y,0}$ +2 mm (COD), which impose strict constraint to the beam injection path. The half aperture of collimators in the ring is designated to ~$12.5\sigma_{x,y,0}$. Simulations of the injection bunch's motion with collimators are carried out to decide the injection efficiency. Top-off injection will be adopted to achieve the higher average luminosity.

## 3 COLLECTIVE EFFECT

In BEPCII, to achieve the high luminosity, the micro-β scheme is adopted. This requires that the bunch length be well controlled to 1.5cm. However, bunch lengthens due to the potential well distortion and microwave instability. Normally, we chose the design bunch under the threshold of microwave instability. This corresponds to an impedance threshold of 0.97 Ω at the design current of 9.8mA. Calculation and beam measurement have shows that the broadband impedance of the present BEPC storage ring is about 4 Ω, thus a strict impedance budget has been made with efforts on designing and optimizing the main vacuum components to reduce the impedance. The main impedance generating elements and their impedance are listed in Table 2. The budget of $(Z/n)_0$ is estimated as 0.23 Ω.

Table 2: Impedance and loss factor of each component

| Component | No. | L (nH) | $k_l$ (V/pC) | HOM (kW) |
|---|---|---|---|---|
| SRF | 1 | | 0.67 | 4.74 |
| Resist. Wall | | | 0.11 | 0.78 |
| BPMs | 68 | 3.3 | 0.08 | 0.57 |
| Bellows | 67 | 0.48 | 0.02 | 0.14 |
| Flange | 200 | 3.0 | 0.003 | 0.02 |
| Mask | 40 | 2.8 | 0.06 | 0.42 |
| Pump. ports | | 0.5 | | |
| Taper | 8 | 4.4 | 0.005 | 0.35 |
| Injection port | 1 | 0.17 | 0.005 | 0.05 |
| Inj. kicker | 2 | 0.8 | 0.04 | 0.28 |
| X-cross | 1 | 0.8 | 0.03 | 0.21 |
| Y-shape | 2 | 2.2 | 0.19 | 1.34 |
| IR | 1 | 0.8 | 0.01 | 0.07 |
| Collimator | 3 | 3.81 | 0.06 | 0.42 |
| Total | | 28.9 | 1.76 | 12.47 |

The longitudinal effective impedance is calculated according to the Heifets-Bane broadband impedance model, and the threshold bunch current for the microwave instability is around 36 mA, which has the reasonable margin. At 9.8 mA, the bunch lengthens by about 5%.

The coupled bunch instabilities can arise from the high-Q resonant structures, such as the RF cavities and the resistive-wall impedance of the beam pipe. In BEPCII, superconducting cavities (SC) are adopted, so HOMs can be well damped. The up bound of the growth rates is firstly estimated by assuming symmetric filling pattern with 99 equally spaced bunches and each bunch has currents of 9.8 mA. This is confirmed by the multi-particle tracking results. The growth time of the fastest growing instability modes is at the same level of the synchrotron radiation damping time, so longitudinal feedback system will be considered. The growth rate of the most dangerous mode due to resistive-wall is of 4.3 ms, so this should be damped with bunch-to-bunch transverse feedback system.

To avoid ion trapping, a clearing gap with 5% of the total buckets absent is required, i.e. with one bunch train of 93 bunches. The Fast Beam-Ion Instability (FBII) has been studied with analytical formulae as well as tracking code, the growth time of FBII is about 3 ms which should be damped with the feedback system.

In analogy to other two ring colliders such as , KEKB and PEPII, the electron cloud instability (ECI) may lead to the beam-size blow-up and luminosity degradation in BEPCII. With the existing formula [4], the threshold electron cloud density leading to transverse mode-coupled instability is higher than that of two B-factories. This may attribute to the smaller circumference of BEPCII. To guarantee the beam performance against ECI, antechamber with the inner surface of beam chamber TiN coated is used in the arc to reduce the primary and secondary electron yields, and in the straight section space may be reserved to wind solenoids to suppress the concentration of electrons near the beam axis. Besides, we are also investigating the possibility to install clearing electrodes to sweep out electrons. A simulation code[5] has been developed to estimate the electron cloud density taking into account each effect due to the antechamber structure, the TiN coating and clearing electrode, respectively. A typical distribution of electron cloud density in the antechamber is shown in Fig.5.

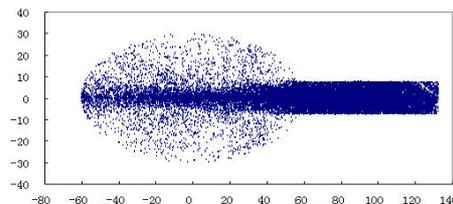

Figure 5: The electron cloud density in the antechamber.

Simulation shows that with antechamber and TiN coating, the EC density is substantially lower than the instability threshold, and feedback system is still needed to overcome the coupled bunch instabilities.

In BEPCII, the beam lifetime is mainly determined by beam-beam bremsstrahlung during beam collision, which is calculated as 5.1 hours at the peak luminosity. Other limits are from the beam-gas scattering which gives beam lifetime of about 26 hrs at the vacuum pressure of $8\times10^{-9}$ Torr, and the Touschek effect of about 7.1 hrs. So the total beam lifetime is around 3.0 hrs. With the top-off injection the maximum average luminosity is expected as 60% of the peak luminosity

## 4 SUMMARY

The main issues in BEPCII accelerator physics design are described. The beam-beam simulation shows the crossing angle of 2×11mrad is acceptable and the beam-beam parameter of 0.04 is reasonable and reachable. The geometry satisfies the requirements for both colliding and synchrotron radiation mode. The IR design realizes the rapid separation of two beams and can meet all the requirements from hardwares. The optics for the collision mode is very flexible on emittance and tunes, with a good dynamic aperture even for $10\sigma_\varepsilon$, off-momentum particles and with multipoles' error as well as the misaliagnment effects. The injection design satisfies the requirements for the top-off injection scheme. For collective effects, a strict impedance budget is made and there is no show stoppers. However, feedback systems are required.